# Scientific Machine Learning Seismology

Tomohisa Okazaki

RIKEN Center for Advanced Intelligence Project

**Abstract**

Scientific machine learning (SciML) is an interdisciplinary research field that integrates machine learning, particularly deep learning, with physics theory to understand and predict complex natural phenomena. By incorporating physical knowledge, SciML reduces the dependency on observational data, which is often limited in the natural sciences. In this article, the fundamental concepts of SciML, its applications in seismology, and prospects are described. Specifically, two popular methods are mainly discussed: physics-informed neural networks (PINNs) and neural operators (NOs). PINNs can address both forward and inverse problems by incorporating governing laws into the loss functions. The use of PINNs is expanding into areas such as simultaneous solutions of differential equations, inference in underdetermined systems, and regularization based on physics. These research directions would broaden the scope of deep learning in natural sciences. NOs are models designed for operator learning, which deals with relationships between infinite-dimensional spaces—a common feature in physics problems such as general solutions to differential equations. NOs show promise in modeling the time evolution of complex systems based on observational or simulation data. Since large amounts of data are often required, combining NOs with physics-informed learning holds significant potential. Finally, SciML is considered from a broader perspective beyond deep learning: statistical (or mathematical) frameworks that integrate observational data with physical principles to model natural phenomena. In seismology, mathematically rigorous Bayesian statistics has been developed over the past decades, whereas more flexible and scalable deep learning has only emerged recently. Both approaches can be considered as part of SciML in a broad sense. Theoretical and practical insights in both directions would advance SciML methodologies and thereby deepen our understanding of earthquake phenomena.

**§1. Introduction**

Machine learning, particularly deep learning, is widely adopted in various fields of natural sciences. In seismology, deep learning has been applied in several stages of earthquake catalog creation, including the phase picking and polarity determination of seismic waves [Mousavi and Beroza (2022), Kubo et al. (2024)]. Such technologies have been realized by training large neural network (NN) models on "big data" such as observational seismic records.

However, it is difficult for many phenomena to obtain sufficient observational data. Ground motions and crustal deformation are typical examples, because their driving forces such as faults exist in the Earth, while observations are mainly limited to the ground surface and the seafloor, and because large earthquakes that cause significant damage rarely occur. In such phenomena, the benefits from deep learning based on big data analysis have been limited.

In natural sciences, universal physical laws derived from phenomena have enabled highly accurate inference through numerical simulations and other methods. Additionally, data analysis techniques have been developed to describe and predict the state and temporal evolution of systems, including their uncertainties, using limited observations. Based on a solid mathematical basis, these methods have greatly advanced our understanding of natural phenomena. However, challenges remain in analyzing large-scale systems and those for which only incomplete physical laws are available.

Since the late 2010s, a research trend has flourished that integrates scientific knowledge into machine learning methods, such as deep learning, to analyze physical phenomena with limited data. This field is collectively referred to as scientific machine learning (SciML) [Baker et al. (2019)]. By combining the strengths of scientific theory and machine learning while compensating for their respective weaknesses, SciML aims to enable large-scale and efficient numerical analysis, the discovery and refinement of theoretical models, and the expansion of target phenomena for inference and prediction. By leveraging the flexibility of deep learning modeling, various approaches that are distinct from traditional methods have been proposed.

In seismology, the amount and characteristics of observational data, such as noise properties and spatiotemporal localization, vary widely. In addition to differential equations, various relationships are known as physical laws, such as constitutive laws based on experiments and scaling laws derived from the statistical properties of earthquakes, each with different forms and degrees of certainty. SciML has the potential to flexibly integrate these diverse observational data and physical laws. By offering new insights into previously unresolved problems, SciML may contribute to a deeper understanding of earthquake phenomena.

In this article, I explain the main methods of SciML and their applications in seismology, including the author's own research. In Chapter 2, I classify and organize the target problems of SciML from multiple perspectives. Chapter 3 provides a detailed explanation of physics-informed learning, which is the core concept of SciML, followed by an overview of operator learning in Chapter 4. These two chapters cover the latest methods based on deep learning. In Chapter 5, we revisit traditional analysis methods through the lens of SciML and offer a broader outlook on SciML.

Table 1. Basic properties of scientific machine learning methods

| Property | Physics-informed neural network (PINN) | Deep operator network (DeepONet) | Fourier neural operator (FNO) |
|---|---|---|---|
| Target quantity | Function | Operator | Operator |
| Input variable | Continuous | Discrete (Fixed) | Discrete (Variable) |
| Output variable | Continuous | Continuous | Discrete (Variable) |
| Input–output space | Arbitrary | Arbitrary | Identical |
| Model geometry | Arbitrary | Arbitrary | Uniform grids |
| Network architecture | Simple | Simple | Complex |
| Key concept | Physics-informed loss function | Universal approximation of operators | Discretization invariance of operators |
| Section in the text | 3.2 | 4.2 (1) | 4.2 (2) |

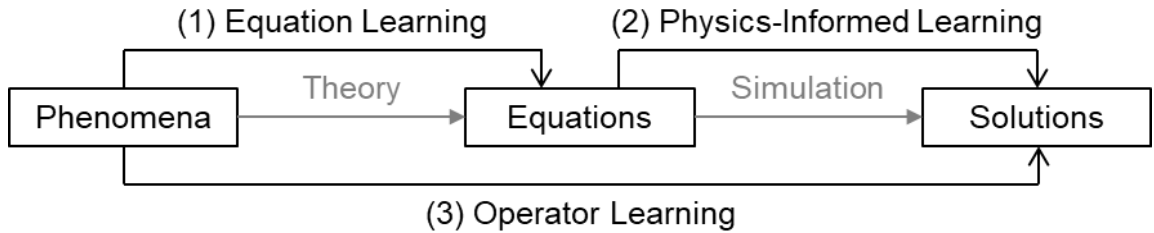

Figure 1. Target problems of scientific machine learning (SciML). Gray and black arrows represent traditional research processes and SciML approaches, respectively.

## §2. The Scope of Scientific Machine Learning (SciML)

SciML refers to a general approach for analyzing natural phenomena through machine learning techniques such as deep learning, and its specific methodologies are diverse. In this chapter, I organize the problem settings and methodologies targeted by SciML from multiple perspectives. Basic properties of representative methods are summarized in Table 1.

**2.1 Modeling Natural Phenomena**

The fundamental goal of natural sciences is to explain and predict various natural phenomena through universal laws. Physical laws are often expressed in equations, with partial differential equations (PDEs) playing a central role. These equations are typically derived from experiments and observations, and solutions are obtained through numerical computations. SciML can be seen as an approach that complements part or all of this process using machine learning techniques (Figure 1).

(1) **Equation Learning**: The use of deep learning for system identification to discover governing equations from observational data has been actively researched since the late 2010s in the machine learning community [Chen et al. (2018), Greydanus et al. (2019)]. Models are designed based on the governing principles of dynamical systems. Their application to seismology has not progressed significantly, and this article does not cover it. While the fundamental equations for seismic wave propagation and crustal deformation are well established, the rheological properties and fault friction laws of rocks are not yet fully understood, leaving room for future development in this area.

(2) **Physics-Informed Learning**: In solving equations, numerical simulations such as the finite difference method and the finite element method have been the dominant approaches. Recently, NN-based PDE solvers, known as physics-informed neural networks (PINNs), were proposed and have garnered significant attention [Raissi et al. (2019)]. Subsequent studies have shown that while PINNs often underperform compared to traditional methods as pure solvers, their usefulness has become evident in other aspects, such as data assimilation from observations. PINNs are one of the key methods that sparked interest in SciML and will be discussed in Chapter 3.

(3) **Operator Learning**: A key feature of deep learning is its ability to learn input–output relationships from abundant data without the need for manually designed features. Therefore, when sufficient observational data are available, solutions can be modeled directly from the phenomena without equations. This is similar to traditional applications, such as seismic phase picking. However, natural phenomena are expressed in continuous variables, and existing machine learning models have typically been applied after discretization (e.g., time variables are discretized at sampling intervals). In SciML, operator learning that aims to model continuous phenomena directly has been actively studied [Li et al. (2021), Lu et al. (2021)], which will be overviewed in Chapter 4.

For operator learning, observational data are typically used as training data. In this article, this is referred to as the observational data-driven approach. However, as mentioned in Chapter 1, sufficient observational data are not always available. If the governing equations of systems are known, the results of numerical simulations can be used as training data. This is referred to the simulation data-driven approach. The latter is generally known as surrogate models, which will be discussed in Section 2.4.1.

**2.2 Incorporating Physical Knowledge**

The primary goal of SciML is the modeling of natural phenomena, but its distinguishing feature lies not merely in the application of existing methods, but in reflecting physical knowledge within machine learning models. This can be seen as introducing biases into statistical models based on prior knowledge on physical systems. Here, these are classified following the pioneering review paper by Karniadakis et al. (2021).

(1) **Observational Bias**: For natural phenomena where sufficient observational data are unavailable, if the governing equations are known, training data can be generated through numerical simulations, allowing the machine learning model to more easily acquire physically consistent input–output relationships. Expanding training data based on physical knowledge is referred to as observational bias.

The machine learning model itself uses a general-purpose structure, and the physical laws are only indirectly reflected through the training data. Therefore, the model may produce unphysical outputs depending on the input values.

(2) **Inductive Bias**: Designing the structure of a machine learning model such that it always satisfies physical laws is referred to as inductive bias. For example, by applying appropriate transformations to the inputs or outputs, symmetry can be maintained, or by using a potential function as the output variable, conservation laws can be ensured. Since this approach strictly enforces the physical laws, it is called a hard constraint.

By strictly adhering to physical laws and limiting the model space, optimization becomes easier, making this an ideal method for incorporating physical knowledge. However, it requires the design of a specialized model structure, which limits its application to relatively simple laws.

(3) **Learning Bias**: Incorporating physical knowledge into the learning algorithm is referred to as learning bias. A typical example is physics-informed learning, where physical equations are expressed as part of the loss function, allowing the integration of complex laws such as PDEs. Since the physical laws are only approximately satisfied, this approach is called a soft constraint.

Compared to inductive bias, although the laws are not strictly enforced, the advantage is that a wider range of laws can be easily incorporated. However, since the learning algorithm tends to become more complex, optimization is often more challenging.

The three types of biases mentioned above are not mutually exclusive, and combining multiple types of knowledge can enhance the model's generalization performance.

**2.3 Classification by Learning Method**

Machine learning makes inferences based on the statistical properties of data and is referred to as a data-driven approach. In contrast, Raissi et al. (2019) introduced physics-informed learning, where NNs are trained by incorporating differential equations, allowing inferences based on physical laws. This represents a groundbreaking idea that fundamentally changes the learning method.

NNs are models that learn relationships between finite-dimensional vectors, that is, "functions." In contrast, operator learning focuses on learning relationships between infinite-dimensional objects, such as functions, known as "operators." Conventional NNs struggle to handle such relationships directly and the construction of specialized models is necessary.

In summary, research has traditionally focused on "data-driven function learning." However, with the recent emergence of concepts like physics-informed learning and operator learning, these approaches have become major areas of interest in SciML research.

Table 2. Deep learning methods for modeling physical systems

| Modeling task | Data-driven surrogate model | Physics-informed solver |
|---|---|---|
| Forward modeling (Single solution) | (Neural network)[1] | Physics-informed neural network[4] |
| Simultaneous solution (Finite dimension) | Neural network[2] | Physics-informed neural network[5] |
| Operator learning (Infinite dimension) | Neural operator[3] | Physics-informed neural operator[6] |

Superscripts correspond to enumeration in Section 2.4.1.

## 2.4 Modeling Types

### 2.4.1 Forward Analysis

Even before the advent of SciML, surrogate models have been developed to interpolate the results of forward calculations in the parameter space using NNs. Numerical simulations are typically carried out by specifying parameters such as initial conditions, material properties, and geometric configurations, requiring repeated computations for each set of conditions. High-resolution simulations can demand significant computational resources and time (e.g., hours to days for a single analysis), making multiple iterations or real-time analysis difficult. The primary motivation for using NNs is their ability to infer results at high speed (e.g., in less than a second). This corresponds to the simulation data-driven approach in Section 2.1 and complete observational bias (i.e., no observational data) in Section 2.2.

Specifically, by expressing the solution $u(x;a)$ for condition $a$, an NN is trained on the numerical simulation results $(u(x;a_1),…,u(x;a_N))$ for multiple conditions $(a_1,…,a_N)$. Once trained, the NN can rapidly infer the solution $u(x;a)$ for arbitrary parameter $a$. One of the key advantages is that the NN represents parameters as continuous variables, allowing it to learn the continuous dependencies from discrete computational results. However, training requires numerous numerical simulations and optimization of the NN model, which generally demands substantial computational resources.

The problem settings for surrogate models are shown in Table 2. (1) For single solutions that can be obtained through one numerical computation, surrogate models are rarely used. (2) Surrogate models for finite-dimensional parameters using NNs have been developed [e.g., DeVries et al. (2017)]. In SciML, there is growing interest in (3) operator learning, which involves surrogating infinite-dimensional conditions, such as initial conditions and domain shapes. This has led to active research on neural operators (NOs) [Li et al. (2021), Lu et al. (2021)].

Traditionally, the data-driven approach involved solving equations using numerical simulations and then interpolating them with NNs. Owing to physics-informed learning, the entire process can now be performed using NNs. NNs are no longer merely surrogates for conventional methods; they have evolved into new forward solvers that can solve equations in the parameter space continuously. As shown in Table 2, both (4) single solution and (5) finite-dimensional parameter problems can be handled by PINNs. Originally, PINNs were proposed for solving single solutions [Raissi et al. (2019)], but subsequent research expanded them to handle simultaneous solutions. This latter capability is a key perspective for the practical application of PINNs and will be discussed in Section 3.4.1. Additionally, (6) physics-informed neural operators (PINOs) have also been proposed to address operator learning [Wang et al. (2021), Li et al. (2024b)].

**2.4.2 Inversion Analysis**

Numerical simulations and surrogate models are techniques used to solve physical equations under specified conditions, without explicitly incorporating observational data. In contrast, inversion analysis is a technique used to estimate unknown parameters or conditions in a physical model based on partial observational data. Typically, only discrete and noisy observational data are available for estimating continuous quantities, leading to ill-posed problems where the solution is not uniquely determined. Therefore, methods have been developed to maintain the stability of the solution and to quantify the uncertainty of the estimates.

SciML, which incorporates physical knowledge into data science, is expected to have significant applications in inversion analysis. There are mainly two approaches. One approach is to directly perform inversion analysis using SciML. PINNs are used because they integrate the physical model and observational data into the NN's loss function. While challenges remain in terms of the reliability of solutions and uncertainty quantification, PINNs have shown promise in addressing ill-posed problems [Karniadakis et al. (2021)].

The other approach involves incorporating SciML into traditional inversion methods. In sampling methods such as Markov Chain Monte Carlo, the likelihood must be calculated each time a sample is obtained, which requires fast forward computations. While analytical solutions are often used for this forward calculation, applying SciML techniques such as surrogate models and simultaneous solutions (Table 2) allows for analysis in more complex problem settings. Besides fast inference, NNs have advantages over traditional numerical simulations in providing a differentiable representation of physical variables, enabling efficient parameter optimization [Azizzadenesheli et al. (2024)].

**§3. Physics-Informed Learning**

**3.1 Introduction**

Many natural phenomena can be described by systems of equations, and by solving these equations under given conditions, the behavior of the physical system can be explained and predicted. In general, analytical solutions are only available for simple problem settings, while more complex phenomena require numerical computation. Let us consider the task of solving the following simultaneous equation numerically:

$$\begin{cases} x + y = 5 \\ x - y = 1 \end{cases} \quad (1)$$

To solve this, we define a loss function based on the squared sum of residuals from the equations:

$$L(x, y) = (x + y - 5)^2 + (x - y - 1)^2 \quad (2)$$

The loss function $L$ is non-negative and equals zero if and only if equation (1) is satisfied. In other words, solving the simultaneous equation (1) is equivalent to minimizing the loss function (2). The minimum value can be searched for numerically using gradient descent methods.

Next, let us consider solving the system of differential equations:

$$\begin{cases} \frac{du}{dx} = u(x) \\ u(0) = 1 \end{cases} \quad (3)$$

over the interval $0 \leq x \leq 1$. Similar to the previous example, we define the loss function as:

$$L[u] = \int_0^1 \left(\frac{du}{dx} - u(x)\right)^2 dx + (u(0) - 1)^2 \quad (4)$$

Here, because $u(x)$ is a function, $L$ becomes a functional, and since the differential equation is defined over an interval, an integral appears in the loss function. Solving the differential equation (3) is equivalent to minimizing the loss function (4). By transforming the system of equations into an equivalent optimization problem in this way, machine learning techniques can be applied. Since differential equations primarily describe physical laws, this learning approach is called physics-informed learning.

However, minimizing the functional $L$ is a variational problem over the function $u(x)$, which is infinite-dimensional and challenging to execute on a computer. To address this, we restrict the the function $u(x)$ to a finite-dimensional space by representing it as $u(x;\theta)$ with parameter $\theta$. By doing so, the functional (4) is reinterpreted as a function $L(\theta)$ of the parameters $\theta$, and minimization can be performed accordingly. Examples of parametric functions include linear combinations $u(x;\theta) = \sum_{i=1}^M \theta_i \phi_i(x)$ of basis functions $(\phi_1(x), \ldots, \phi_M(x))$. Those with NNs are referred to as physics-informed neural networks (PINNs) discussed in detail below.

### 3.2 Physics-Informed Neural Network (PINN)
### 3.2.1 Forward Modeling

PINNs are numerical solution methods for differential equations by performing physics-informed learning using NNs [Raissi et al. (2019)]. The unknown solution is represented as $u^{NN}(x;\theta)$ using the NN, and the loss function is defined by the residuals from the system of differential equation:

$$L = L_{\text{phys}} = L_{\text{PDE}} + L_{\text{BC}} + L_{\text{IC}} \quad (5)$$

Here, $L_{\text{phys}}$ represents the loss based on the physical laws, with $L_{\text{PDE}}$, $L_{\text{BC}}$, $L_{\text{IC}}$ denoting the residuals from PDE, boundary condition (BC), and initial condition (IC), respectively. By training the NN using gradient descent, the optimal parameters $\theta^*$ are found, and the resulting $u^{NN}(x;\theta^*)$ becomes the approximate solution to the differential equation.

Because the method deals with differential equations, the loss function includes derivatives of $u^{NN}(x;\theta)$ with respect to $x$ (e.g., $du/dx$ in equation 4). These derivatives of NNs are computed exactly and efficiently using an algorithm called automatic differentiation [Baydin et al. (2018)]. Automatic differentiation is an essential technique for implementing PINNs. The basic structure of a PINN is illustrated in Figure 2.

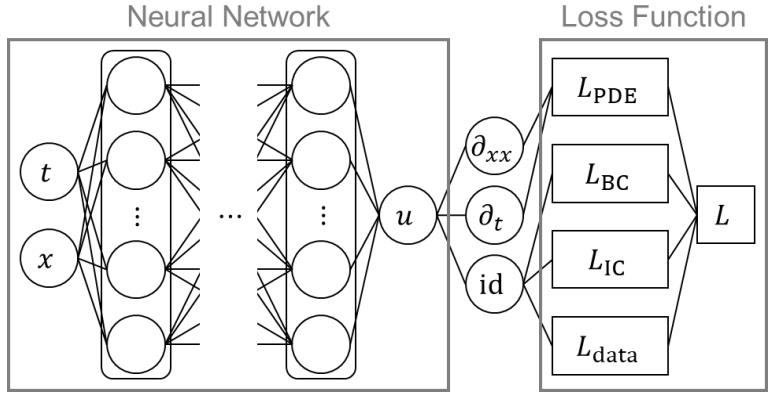

Figure 2. Basic structure of physics-informed neural networks. The diffusion equation (equation 6) is shown as an example. The symbols $\partial$ and id represent partial differential operators (with respect to the subscript variables) and the identity map, respectively. PDE, partial differential equation; BC, boundary condition; IC, initial condition.

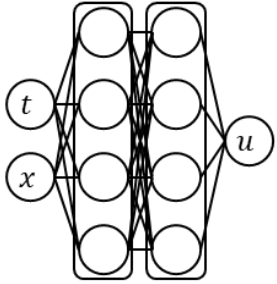 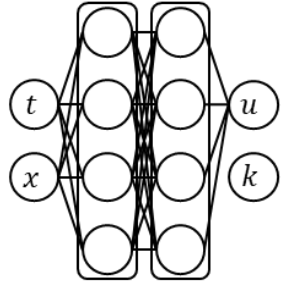 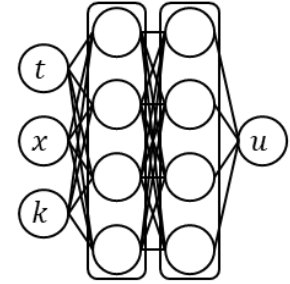

Figure 3. Modeling approaches with physics-informed neural networks. Top panels represent the architecture of neural networks. Bottom panels represent the input domains. Cross marks indicate collocation points (white, partial differential equation; light gray, boundary condition; heavy gray, initial condition) and triangles indicate observational data.

The loss function is abstractly represented as an integral over the domain where the PDE, BC, and IC are defined, as shown in equation (4). In practical computations, this integral is approximated by the sum of values at a finite number of points called collocation points. There are two main approaches for selecting collocation points. The first approach is to place many points on a grid and reuse them throughout the training process. The second approach involves sampling the collocation points according to a specified probability distribution at each iteration of the training. Both methods are commonly used [Das and Tesfamariam (2022), Wu et al. (2023)].

As an example, let us consider the following diffusion equation using a PINN (Figure 3a):

$$u_t = k u_{xx} \tag{6a}$$

$$u(t, 0) = u(t, L) \tag{6b}$$

$$u(0, x) = f(x) \tag{6c}$$

where the subscripts denote partial derivatives with respect to the corresponding variables. The value of the diffusion coefficient $k$ is specified. Equation (6b) represents periodic boundary conditions, and $f(x)$ is the given initial condition. The solution is expressed using an NN as $u^{NN}(t, x; \theta)$, and the loss function is defined as:

$$L = L_{\text{PDE}} + L_{\text{BC}} + L_{\text{IC}} \tag{7a}$$

$$L_{\text{PDE}} = \frac{1}{N_{\text{PDE}}} \sum_{i=1}^{N_{\text{PDE}}} \left( u_t^{NN}(t_i^{\text{PDE}}, x_i^{\text{PDE}}; \theta) - k u_{xx}^{NN}(t_i^{\text{PDE}}, x_i^{\text{PDE}}; \theta) \right)^2 \tag{7b}$$

$$L_{\text{BC}} = \frac{1}{N_{\text{BC}}} \sum_{i=1}^{N_{\text{BC}}} \left( u^{NN}(t_i^{\text{BC}}, 0; \theta) - u^{NN}(t_i^{\text{BC}}, L; \theta) \right)^2 \tag{7c}$$

$$L_{\text{IC}} = \frac{1}{N_{\text{IC}}} \sum_{i=1}^{N_{\text{IC}}} \left( u^{NN}(0, x_i^{\text{IC}}; \theta) - f(x_i^{\text{IC}}) \right)^2 \tag{7d}$$

The terms $L_{\text{PDE}}$, $L_{\text{BC}}$, and $L_{\text{IC}}$ are defined as the root-mean-square residuals over collocation points $\{t_i^{\text{PDE}}, x_i^{\text{PDE}}\}_{i=1}^{N_{\text{PDE}}}$, $\{t_i^{\text{BC}}\}_{i=1}^{N_{\text{BC}}}$, and $\{x_i^{\text{IC}}\}_{i=1}^{N_{\text{IC}}}$ selected from the domains of the PDE, BC, and IC, respectively.

In equation (7b), $u_t^{NN}$ and $u_{xx}^{NN}$ represent the derivatives of the NN output with respect to the input variables. These derivatives are computed efficiently using automatic differentiation. By optimizing the NN parameters $\theta$ to minimize the loss function $L$ in equation (7), the approximate solution $u^{NN}(t, x; \theta^*)$ to the system of equations (6) can be obtained.

NNs have been used for supervised learning, where they learn the input–output relationships from training data. To achieve high performance, a large amount of data are required, leading to the rise of data-driven approaches and big data analysis, where the quantity and quality of the dataset are paramount. In contrast, PINNs can solve equations without relying on training data, functioning similarly to numerical simulations. In other words, PINNs free deep learning from the need for big data, expanding the range of applications in natural sciences.

NNs are highly flexible function approximators, and the conditions to be satisfied are expressed through the loss function. If the loss function represents the residuals between the data and the model's predictions, the NN addresses supervised learning. If the loss function represents the residuals from a differential equation, the NN provides an approximate solution to the differential equation. This might seem obvious in hindsight, but it had been overlooked, much like the concept of "Columbus' egg." The significance of Raissi et al. (2019) lies in materializing this idea in a clear way by using the computational technique of automatic differentiation.

**3.2.2 Inverse Modeling and Data Assimilation**

When, in addition to physical laws, the values of the solution at discrete points are known through observations, a data residual term $L_{\text{data}}$ can be added to the loss function:

$$L = L_{\text{phys}} + L_{\text{data}} \tag{8}$$

By minimizing this loss function, a solution satisfying both the physics and data can be obtained. This can be seen as either making the optimization easier by adding data to the physics-informed learning ($L = L_{\text{phys}}$), or as stabilizing the inference by incorporating physical knowledge into a supervised learning ($L = L_{\text{data}}$).

A particularly interesting case arises when there are unknown parameters in the system of equations. For example, suppose that the diffusion coefficient $k$ in the diffusion equation (6) is unknown, and we seek to estimate its value from observational data (Figure 3b). By denoting the observational data as $\{(t_i^{\text{data}}, x_i^{\text{data}}, u_i^{\text{data}})\}_{i=1}^{N_{\text{data}}}$, the loss function is defined as:

$$L = L_{\text{PDE}} + L_{\text{BC}} + L_{\text{IC}} + L_{\text{data}} \tag{9a}$$

$$L_{\text{data}} = \frac{1}{N_{\text{data}}} \sum_{i=1}^{N_{\text{data}}} \left(u^{\text{NN}}(t_i^{\text{data}}, x_i^{\text{data}}; \theta) - u_i^{\text{data}}\right)^2 \tag{9b}$$

Here, $L_{\text{PDE}}$, $L_{\text{BC}}$, and $L_{\text{IC}}$ are the same as in the forward analysis (equation 7), but with $k$ now treated as an estimated parameter rather than a given value. By optimizing both the NN parameters and the estimate for $k$, represented as $(\theta, k)$, to minimize the loss function (9), the estimate of diffusion coefficient $k^*$ and the corresponding solution $u^{\text{NN}}(t, x; \theta^*)$ can be obtained simultaneously. If the diffusion coefficient $k(x)$ depends on space, the problem can be handled by constructing two separate NNs: one for the solution $u^{\text{NN}}(t, x; \theta_u)$ and the other for the spatially dependent diffusion coefficient $k^{\text{NN}}(x; \theta_k)$. The parameters $(\theta_u, \theta_k)$ are optimized simultaneously to minimize the loss function. Moreover, if the observational data is available only for a limited time period $t \leq T$, by placing collocation points for times $t > T$ as well, time evolution can be predicted through data assimilation.

The fact that PINNs can be easily extended to inverse modeling was demonstrated in the original paper by Raissi et al. (2019). A notable feature is that this can be achieved simply by modifying the structure of the NN and the loss function. Traditionally, methods for forward analysis (e.g., analytical solutions or numerical simulations) and inversion analysis (e.g., least squares methods or Bayesian statistics) had to be appropriately combined. The ability to implement both forward and inverse modeling within a unified framework is one of the key factors why PINNs have attracted significant attention across various fields of science, including seismology.

**3.2.3 Basic Characteristics**

PINNs were proposed as a deep learning-based approach for solving both forward and inverse problems in physical systems, and they possess features that distinguish them from traditional methods. A significant merit is their ease of implementation. Unlike numerical simulations, PINNs do not require discretization of the model domain or equations. Instead, PDEs can be solved by directly expressing the equations as a loss function. This simplicity and ease of implementation have been major factors in the rapid and widespread adoption of PINNs. Moreover, their continuous representation and differentiability are suitable for analysis in complex geometries and parameter optimization, respectively.

However, several challenges have emerged when using PINNs as solvers. One major issue is that NN parameters are generally optimized stochastically, so there is no guarantee of convergence within a practical amount of time. Additionally, the theoretical framework that provides the value of loss function to ensure a specific level of accuracy in the solution is still in progress [De Ryck and Mishra (2024)]. These challenges pertain to both computational time and accuracy. Furthermore, difficulties in optimization have been noted, such as the convergence to trivial local minima and ignoring causality in dynamical systems [e.g., Krishnapriyan et al. (2021), Wang et al. (2024)]. On the technical side, optimal design of the NN, including aspects like the network architecture, the relative weights of the terms in the loss function, the selection of collocation points, parameter initialization, and choice of optimization functions, remains an open question. Many studies are actively exploring these areas to improve the performance of PINNs [e.g., Müller and Zeinhofer (2023), Wang et al. (2023)].

Compared to data-driven approaches, PINNs enhance the generalization performance on rare or unknown events and improve data efficiency by incorporating physical knowledge, helping to suppress unnatural solutions. This would broaden the scope of deep learning, which has primarily been limited to big data analysis.

In Earth sciences, PINNs are seen as promising due to their ability to formulate inverse problems in a straightforward manner. However, many geophysical datasets are sparse and noisy, making it crucial to assess the uncertainty of the estimates. Bayesian inference has been introduced into PINNs to address this, but simple methods can handle toy models only, and there are challenges in terms of evaluation performance and computational time for real-world problems. There is a need for the development of practical and efficient approximate Bayesian methods, which will be discussed in Section 5.3.1.

**3.3 Applications in Seismology**

There are numerous studies on PINNs in seismology. In this section, we will focus on pioneering or notable examples, rather than providing a comprehensive list.

**(1) Traveltime**

Traveltime $T(x, x_s)$ refers to the time for seismic waves to propagate from the earthquake source $x_s$ to an observation point $x$. It provides fundamental information in seismology used in tasks such as hypocenter and subsurface structure estimations. Because the seismic wave velocity $V(x)$ depends on the physical properties of the Earth's interior, causing reflection and refraction along the path, the relationship between traveltime and seismic wave velocity is nonlinear. This is described by the eikonal equation:

$$|\nabla T(x, x_s)|^2 = \frac{1}{V(x)^2} \quad (10)$$

where $\nabla$ represents the gradient with respect to the observation point $x$.

First, forward calculation of the traveltime $T(x, x_s)$ was performed with a known velocity field $V(x)$ [Smith et al. (2020), Waheed et al. (2021b)]. Incorporating theoretical properties, such as reciprocity, by designing the input and output layers of the NN (i.e., inductive bias in Section 2.2) was shown to significantly improve the optimization efficiency of PINNs [Grubas et al. (2023)].

Furthermore, traveltime tomography has been conducted to estimate the velocity structure from observed travel times. By using two NNs—one for the traveltime $T(x, x_s; \theta_t)$ and the other for the velocity $V(x; \theta_v)$—and optimizing them simultaneously, an estimate for the velocity $V(x; \theta_v^*)$ can be obtained [Waheed et al. (2021a)]. Applications to real-world data have also been performed [Chen et al. (2022)], and uncertainty quantification using Bayesian inference has been explored [Agata et al. (2023), Gou et al. (2023)]. Application of PINNs has advanced the most among the various fields in seismology.

**(2) Wave Propagation**

PINNs have seen rapid development in fluid dynamics [Raissi et al. (2020), Cai et al. (2021)], and wave propagation is a typical subject of analysis in seismology as well. Initially, simpler acoustic (scalar) waves were analyzed [Moseley et al. (2020), Rasht-Behesht et al. (2022)]. Many studies have also focused on transforming the problem into the frequency domain to solve the Helmholtz equation [Song et al. (2021), Alkhalifah et al. (2021)]. This approach simplifies the problem because the equations are independent of frequency, thereby reducing the number of input variables for the PINN and making the optimization easier. Elastic waves are addressed by Ren et al. (2024). Inverse analysis for wave propagation has also been attempted, but it remains mostly in the stage of synthetic tests [Song and Alkhalifah (2021), Rasht-Behesht et al. (2022)]. In these inverse problems, two NNs—one for the wavefield and the other for the velocity structure—are optimized simultaneously. At present, there are only limited examples of PINN applications to tsunamis [Leiteritz et al. (2021)].

**(3) Fault and Crustal Deformation**

Following developments in traveltime and wave propagation, applications to finite faults have begun. Okazaki et al. (2022b) analyzed crustal deformation resulting from earthquakes. Specifically, forward modeling of dislocation models that describe crustal deformation caused by fault and plate motion are carried out, by assuming two-dimensional antiplane strains, which are commonly used to model strike-slip faults. To address challenges such as displacement discontinuities along the fault and stress singularities near the fault tips, the polar coordinate system was introduced. If discontinuities in elastic properties exist, multiple NNs were used by imposing compatibility conditions on the contact surfaces. This approach enabled accurate analysis of deformation in complex subsurface structures, surface topography, fault geometry, and slip distribution, even near faults and material discontinuities. This approach was extended to two-dimensional inplane strains, which are commonly used to model dip-slip faults, and invere modeling [Okazaki et al. (2025)].

Analysis of fault motion based on the rate-and-state dependent friction law began with the spring–slider model [Fukushima et al. (2023)]. The study focused on slow earthquakes and included not only forward and inverse analyses but also data assimilation, which predicted time evolution from a limited period of observational data. Subsequent studies performed analysis in two-dimensional antiplane strains [Rucker and Erickson (2024)] and three-dimensional structures [Fukushima et al. (2024)]. As PINNs continue to be applied to a wider variety of earthquake phenomena, they are expected to further penetrate the field of seismology.

**3.4 New Developments**

PINNs were initially proposed as a method for solving PDEs [Raissi et al. (2019)]. The concept of physics-informed learning, which incorporates physical laws into the loss function, is highly general and flexible, leading to several extensions in subsequent research. In this section, we focus on developments related to seismology, particularly the generalization of forward and inverse analyses through simultaneous solution methods, as well as the analysis of underdetermined systems and physical regularization.

**3.4.1 Simultaneous Solutions**

PINN forward modeling gives solutions under given conditions. However, when multiple conditions are analyzed, PINNs must be trained on individual conditions, resulting in a significant computational cost. The effectiveness of transfer learning—using the solution obtained under one condition as the initial value for another condition to accelerate convergence—has been demonstrated in several studies [e.g., Krishnapriyan et al. (2021), Okazaki et al. (2022b)]. However, this does not lead to a fundamental reduction in computational time. Currently, in many cases, PINNs lag behind traditional numerical simulations in terms of both accuracy and time, limiting their utility as a forward analysis tool.

Solutions for arbitrary conditions can be obtained simultaneously by leveraging the structure of NNs (Table 2). Taking the diffusion equation as an example, in forward modeling, the diffusion coefficient $k$ is treated as a known constant. Therefore, a single training yields the solution for the given value of $k$, and retraining is required for different values (Figure 3a). In contrast, the simultaneous solution method treats the diffusion coefficient $k$ as an additional input variable to the NN, alongside the spatiotemporal variables (Figure 3c). The loss function remains the same as in forward modeling (equation 7), with the only difference being that $k$ is no longer a constant but an input to the NN. Although training time generally increases due to the additional input variable, subsequent inference can be performed quickly. Capturing the continuous dependence on the parameter is a significant benefit of NNs.

In seismology, the eikonal equation is a typical example of simultaneous solutions. Equation (10) is a differential equation with respect to the observation point $x$, and in finite difference methods, the source $x_s$ is fixed to compute the traveltime. In contrast, PINNs can solve the traveltime $T(x, x_s)$ as a function of both $x$ and $x_s$ simultaneously. This feature is particularly powerful in hypocenter estimation. When iteratively optimizing the source location $x_s$, it is necessary to repeatedly compute the traveltime from each updated source $x_s$ to every observation point $x$. In finite difference methods, this requires solving the equation each time $x_s$ is updated. In contrast, once a PINN is trained for a given subsurface structure, it can quickly infer traveltimes for any $x_s$, enabling efficient hypocenter estimation. Smith et al. (2022) demonstrated its effectiveness clearly by using a PINN for forward traveltime computations in hypocenter estimation based on Bayesian inference.

As described above, PINNs can perform simultaneous solutions by treating the parameters specifying the PDE conditions as input variables. However, the inputs are limited to finite-dimensional parameters, such as the diffusion coefficient (a scalar) and source location (a vector). Infinite-dimensional objects, such as conditions or spatially varying initial PDE coefficients, cannot be directly used as input variables. These infinite-dimensional objects must be represented by finite-dimensional parametric functions [Sun et al. (2020), Ren et al. (2024)], but this introduces a trade-off between expressiveness and computational cost. Operator learning discussed in the next chapter is a field that attempts to directly model relationships in infinite-dimensional spaces.

Okazaki et al. (2024) achieved simultaneous solutions to the problem of operator learning using PINNs by incorporating domain knowledge. The simultaneous solution of crustal deformation for variable fault geometries (shapes) and slip distributions (functions) involves handling infinite-dimensional parameters, which cannot be directly addressed by PINNs (Table 2). In antiplane strains, Okazaki et al. (2024) introduced the concept of fault geometry invariance that if fault shapes share common endpoints, the resulting crustal deformation due to uniform slip is essentially the same, regardless of the specific geometry of the fault. Based on this insight, a dislocation potential—which represents the displacement field arising from uniform slip on a linear fault that connects a reference point to an arbitrary point—was defined and shown to produce crustal deformation due to any fault slips. In essence, the information of a fault surface (an infinite-dimensional curve) can be reconstructed from just the fault endpoints (a two-dimensional point). Using a PINN for solving the dislocation potential simultaneously and applying the principle of superposition in inference realized an efficient crustal deformation analysis due to arbitrary fault geometry and slip distribution.

### 3.4.2 Underdetermined Systems and Physical Regularization

PINNs have been applied under the assumption that the governing equations of physical systems are known, allowing for forward analysis and the estimation of unknown parameters. However, in many natural phenomena, it is not always possible to identify a closed system of equations, owing to the lack of governing laws and the use of approximate constitutive laws based on experiments. Solving such underdetermined systems using numerical simulations or predicting time evolution through data assimilation can be challenging.

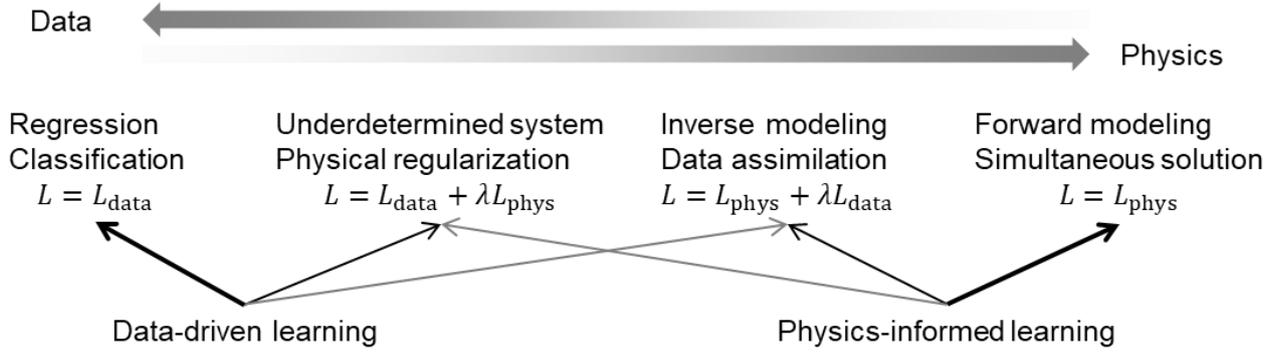

Figure 4. Diverse deep learning approaches for scientific problems. Top arrows represent the amount of available data and physics. $L$ represents the loss function where $L_{\text{data}}$ and $L_{\text{phys}}$ indicate the residuals from data and physics, respectively. $\lambda$ implies that the corresponding term acts as regularization.

In contrast, by treating solving equations as optimization problems, PINNs can provide plausible solutions under the given conditions.

Poulet and Behnoudfar (2023) estimated the spatial distribution of stress, displacement, and elastic constants based on stress orientations and velocity data derived from geological and geodetic observations, respectively, using the equilibrium conditions and stress–strain relationships for elastic bodies. This study highlights the applicability of PINNs to highly underdetermined problems where observational data are sparse and the governing physical equations are insufficient to fully constrain the unknown variables. Their work also extended to the analysis of slip tendency [Poulet and Behnoudfar (2024)]. However, since not all degrees of freedom are constrained by the imposed conditions, there is no guarantee that the solutions obtained are physically accurate, especially in parameter regions with sparse data. Caution must be exercised in interpreting the results.

Borate et al. (2023) constructed an NN to estimate the shear stress and slip velocity of rocks from time histories of ultrasonic wave measurements of velocity and spectral amplitude in laboratory experiments. They enhanced the loss function, which initially consisted of data residuals, by incorporating approximate physical constraints, such as the elastic coupling between the fault and the surrounding, and the ultrasonic transmission coefficient. By introducing these physical regularizations, they demonstrated that even with a limited amount of data, the model achieved high generalization performance. This approach highlights how combining physical knowledge with data-driven models as a form of regularization can stabilize the estimation process. A similar approach was applied to estimate rate-and-state dependent frictional parameters [Borate et al. (2024)].

Methods incorporating specific constraints into NNs based on physical insights has been conducted. For example, consider ground motion prediction equations, which predict ground motion intensities from a few explanatory variables related to earthquakes and observational sites through statistical analysis of past seismic records. NN models have been explored that impose monotonicity as a hard constraint on variables such as magnitude and distance [Okazaki et al. (2021c)] and that learn site-specific conditions using one-hot encoding for each observational station [Okazaki et al. (2021d)]. Although a large amount of seismic data are being accumulated, approaching what can be considered big data, strong motion records from large earthquakes at near-source stations that lead to significant earthquake damage remain rare. In such skewed datasets, physical regularization is expected to play a crucial role.

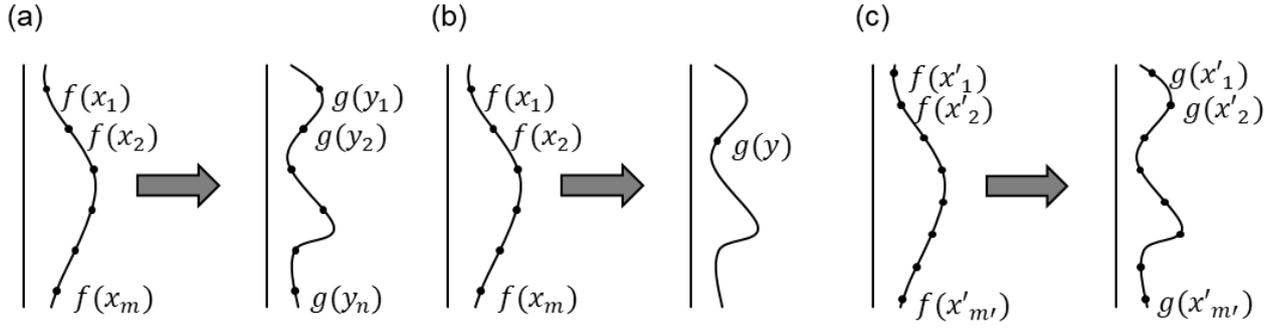

Figure 5. Modeling approaches for operators between continuous functions. (a) Fixed discretization. The locations $(x_1, ..., x_m)$ and $(y_1, ..., y_n)$ are fixed. (b) Fixed discretization for inputs and continuous representation for outputs. (c) Variable discretization for individual input–output pairs. The number $m'$ and positions $(x'_1, ..., x'_{m'})$ can be different for each input.

By viewing physics-informed learning as a regularization for data-driven models, a wide range of physical insights can be incorporated, from precise physical laws such as differential equations, to fundamental principles like symmetry and conservation laws, and even constitutive and empirical laws. A key advantage is the ability to handle relationships involving derivatives through automatic differentiation, as well as the use of approximate relationships due to soft constraints. Crucially, data-driven learning and physics-informed learning are not mutually exclusive but can be integrated to create flexible modeling approaches that adapt to the availability of observational data and the governing physical laws (Figure 4). This fusion allows deep learning to extend beyond big data contexts.

## §4. Operator Learning

### 4.1 Introduction

In natural sciences, problems involving transformations of functions frequently arise (Figure 5). For example, suppose seismic motion at the bedrock as input $f(x)$ and that at the surface as output $g(y)$, where $x$ and $y$ both represent time. The transformation $A: f \mapsto g$ corresponds to the amplification factors of the shallow soil layers. Deep learning has shown great potential for solving such end-to-end problems [e.g., Ross et al. (2018), Zhu and Beroza (2019), Mousavi et al. (2020)]. Typically, to apply deep learning to these problems, the sampling frequency and the length of the seismic records are fixed, allowing the input and output variables to be represented as vectors of fixed dimensions (Figure 5a). In this case, NNs learn a "function" between these vectors.

However, natural phenomena are described using continuous variables (at least in classical theory). In the case of seismic motion, a NN trained on those at 100 Hz sampling cannot process those at 120 Hz sampling. This highlights how discretization introduces an artificial limitation, discarding some of the inherent information in natural phenomena. To address this, approaches that preserve the continuity of variables during modeling have begun to emerge. Since the transformation $A$ can be viewed as an "operator" that maps functions to functions, this approach is referred to as operator learning. While standard machine learning transforms finite-dimensional vectors, operator learning transforms infinite-dimensional functions, making it a qualitatively more advanced problem. Models of operator learning using NNs are called neural operators (NOs) (in a broad sense), which require specialized model architectures.

In fact, handling the output function $g(y)$ continuously is relatively straightforward. By inputting a set of discrete input values and the point of interest for the output as $(f(x_1), \ldots, f(x_m), y)$, the corresponding $g(y)$ can be produced as the output (Figure 5b). Since the output only depends on the specific point $y$, there is no need to consider other points simultaneously. It is known that NNs can approximate operators to any desired accuracy with this structure, which is the idea behind deep operator network (DeepONet) [Lu et al. (2021)]. Because it still operates in a finite-dimensional input space, it can be seen as a partial form of operator learning.

To input $f(x)$ as a continuous function would require infinite-dimensional information, which cannot be directly handled by traditional NNs. To address this, the (narrow sense) NO was developed to discretize the input and output functions in a flexible manner, allowing for essentially treating functions [Li et al. (2021)] (Figure 5c). Inputs and outputs can be not only a specific discretization $(x_1, \ldots, x_m)$ but also arbitrary discretization $(x'_1, \ldots, x'_{m'})$ with different numbers and positions. The key innovation here is discretization invariance (or discretization convergence), which requires that the model converges to a single continuous operator as the discretization becomes finer to ensure the consistency of the model. While the individual inputs and outputs are still finite-dimensional vectors, the space they span are infinite-dimensional, making this approach a more genuine operator learning.

### 4.2 Neural Operator (NO)

The goal of operator learning is to model an operator $G: \mathcal{A} \to \mathcal{U}$ that maps an input function $a \in \mathcal{A}$ ($a: D \to \mathbb{R}^{d_a}$) to an output function $u \in \mathcal{U}$ ($u: D' \to \mathbb{R}^{d_u}$). Typically, in the context of PDE systems, $a(x)$ represents the input conditions, such as initial conditions or PDE coefficients, while $u(y)$ represents the solution, and the goal is to estimate the solution operator. However, the approach is not limited to PDEs and can be applied to general operators. NOs model the operator $G$ using an NN $G^{NN}(a; \theta)$, whose parameters $\theta$ are optimized using a data-driven approach by minimizing the residuals over a training dataset $\{(a_i, u_i)\}_{i=1}^N$:

$$L(\theta) = \frac{1}{N} \sum_{i=1}^{N} \sum_{j=1}^{L} (G^{NN}(a_i; \theta)(y_j) - u_i(y_j))^2 \tag{11}$$

where $(y_1, \ldots, y_L)$ are discrete points at which the loss function is evaluated (Table 2).

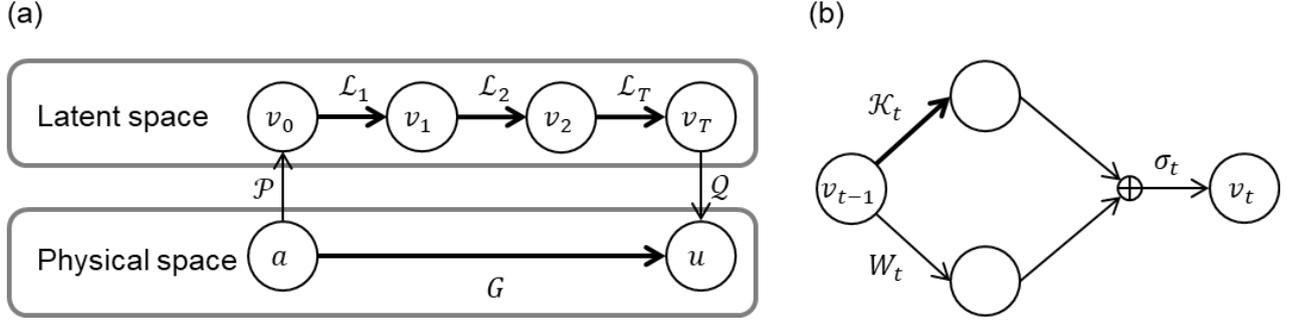

Figure 6. Basic structure of neural operators. (a) A target operator $G$ is composed of a local lifting operator $\mathcal{P}$ that maps input physical functions $a$ into the high-dimensional latent functions, global operators $\mathcal{L}_t$, and a local projection operator $\mathcal{Q}$ that maps latent functions to output physical functions $u$. (b) A global operator $\mathcal{L}_t$ is composed of the sum of an integral operator $\mathcal{K}_t$ and a linear transformation $W_t$ and a nonlinear activation function $\sigma_t$. Bold and thin arrows represent global and local operators, respectively.

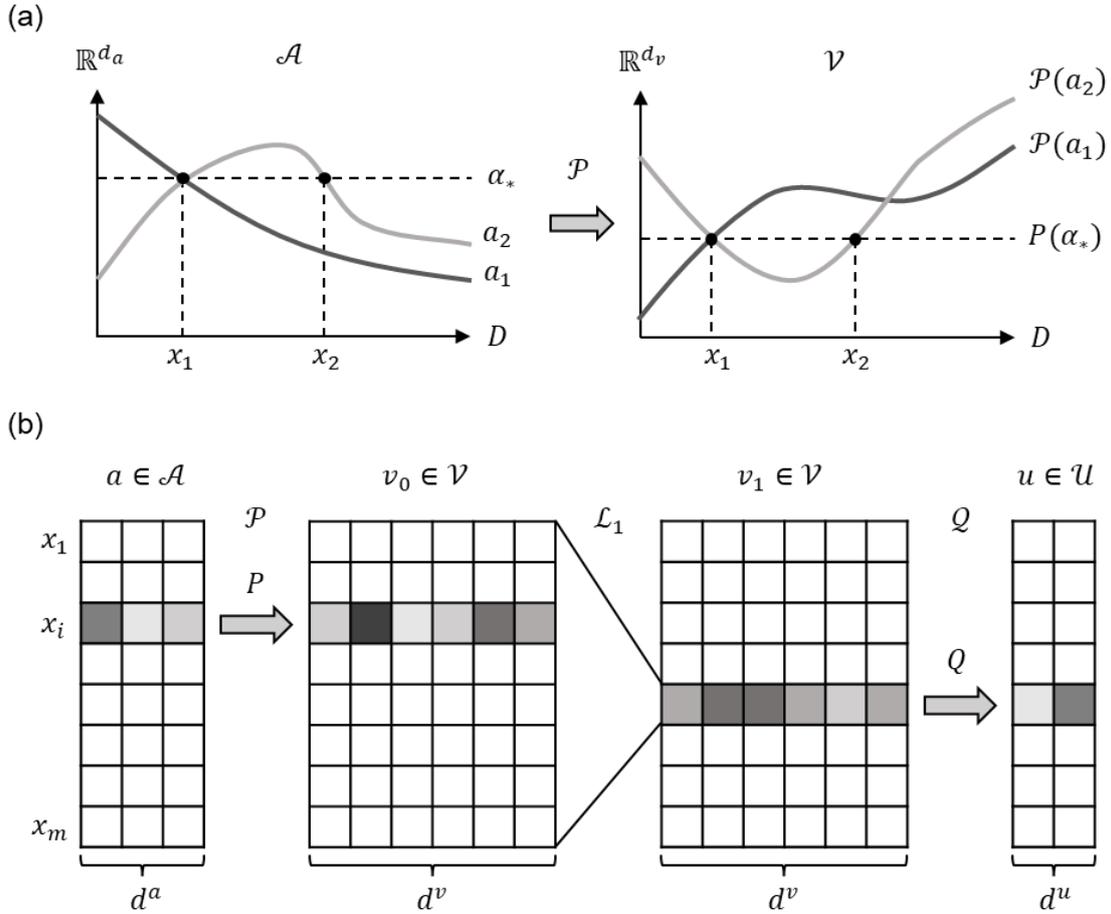

Figure 7. Locality of operators. (a) Property of local operators. If $\mathcal{P}$ is a local operator, $a_1(x_1) = a_2(x_1) = a_2(x_2)$ leads to $(\mathcal{P}(a_1))(x_1) = (\mathcal{P}(a_2))(x_1) = (\mathcal{P}(a_2))(x_2)$. (b) Discrete representation of neural operators. Local operators $\mathcal{P}$ and $\mathcal{Q}$ treat each point $x_i$ separately, whereas a global operator $\mathcal{L}_1$ considers all points including their positions.

**(1) Deep operator network (DeepONet)**

DeepONet, introduced by Lu et al. (2021), is one of the first NN models to handle operator learning. It is based on the construction used in the proof of the universal approximation theorem for operators [Chen and Chen, (1995)]. The operator $G[a]$ is expressed as:

$$G[a](y) = \mathbf{b}[a] \cdot \mathbf{t}(y) \tag{12}$$

where $\mathbf{b}$ and $\mathbf{t}$ are $p$-dimensional vectors (with $p$ arbitrarily specified). By representing the functional $\mathbf{b}[a]$ using the values of $a(x)$ at fixed discrete points $(x_1, \ldots, x_m)$ (Figure 5b) and by employing NNs, the basic structure of DeepONet can be obtained:

$$G[a](y) = \mathbf{b}^{\text{NN}}(a(x_1), \ldots, a(x_m); \theta_b) \cdot \mathbf{t}^{\text{NN}}(y; \theta_t) \tag{13}$$

This can be seen as $p$-dimensional basis function expansion, where $\mathbf{t}^{\text{NN}}(y; \theta_t)$ acts as basis functions and $\mathbf{b}^{\text{NN}}(a(x_1), \ldots, a(x_m); \theta_b)$ represents the expansion coefficients for the input $a(x)$. A key feature is that the basis functions are not predetermined but learned through the NN parameters $\theta_t$. This flexibility allows DeepONet to leverage the high degrees of freedom during training, and once the optimal values $\theta_t^*$ are fixed, robust inference can be performed. DeepONet is relatively easy to implement like PINNs.

**(2) Neural Operator (NO)**

NOs are models that discretize input and output functions but allows the discretization to vary [Li et al. (2021), Kovachki et al. (2023)], although they require that the input and output functions share the same domain ($D = D'$) with a common discretization (Figure 5c). The overall structure of NOs is as follows (Figure 6a):

$$G = \mathcal{Q} \circ \mathcal{L}_T \circ \cdots \circ \mathcal{L}_2 \circ \mathcal{L}_1 \circ \mathcal{P} \tag{14}$$

Here, $\mathcal{P}$ lifts the input function $a(x) \in \mathbb{R}^{d_a}$ from the physical space to a high-dimensional latent space $v_0(x) \in \mathbb{R}^{d_v}$, while $\mathcal{Q}$ projects from the latent space $v_T(x) \in \mathbb{R}^{d_v}$ back to the physical space $u(x) \in \mathbb{R}^{d_u}$. By mapping into high-dimensional space $(d_v > d_a, d_u)$, the expressive power of NNs can be leveraged. The transformations in the latent space, $\mathcal{L}_t: v_{t-1} \mapsto v_t$, performed $T$ times, typically take the form (Figure 6b):

$$v_t = \mathcal{L}_t(v_{t-1}) = \sigma_t(\mathcal{K}_t(v_{t-1}) + W_t v_{t-1}) \tag{15}$$

where $W_t$ is a linear transformation, $\sigma_t$ is a non-linear activation function. $\mathcal{K}_t$ is a global operator discussed later, while $\mathcal{P}$, $\mathcal{Q}$, $W_t$, and $\sigma_t$ are local operators in the following sense: an operator $\mathcal{P}: \mathcal{A} \to \mathcal{V}$ is local if there exists a function $P: \mathbb{R}^{d_a} \to \mathbb{R}^{d_v}$ such that:

$$(\mathcal{P}(a))(x) = P(a(x)) \quad (\forall a \in \mathcal{A}, \forall x \in D) \tag{16}$$

This means that the transformed value $(\mathcal{P}(a))(x)$ at a point $x$ depends only on the value of the function $a(x) \in \mathbb{R}^{d_a}$ at that point, and not on the global shape of the function $a \in \mathcal{A}$ (Figure 7a). In a discrete representation, this is equivalent to forward propagation of an NN $P: \mathbb{R}^{d_a} \to \mathbb{R}^{d_v}$ over a batch of $m$ discrete points (Figure 7b). If the NO were composed solely of local operators, the entire transformation would be reduced to a local operator $G: \mathbb{R}^{d_a} \to \mathbb{R}^{d_u}$, which would simply apply a standard NN at each point $x \in D$.

Thus, the unique global operator, $\mathcal{K}_t$, plays a crucial role in NOs. Generally, it is expressed as the following linear integral operator:

$$(\mathcal{K}_t(v_{t-1}))(x) = \int_D \kappa(x,y) v_{t-1}(y)\, dy \tag{17}$$

By integrating the function $v_{t-1}(y)$ over the domain $D$, the operator accounts for the global behavior of the function, considering the interaction between points across the domain. In the discrete representation, this corresponds to utilizing the values $a(x_i)$ at all points $x_i$, including their positional information (Figure 7b). The kernel $\kappa(x,y)$, which defines $\mathcal{K}_t$, are the target of the modeling. Since $\mathcal{K}_t$ represents an infinite-dimensional transformation, the key challenge is how to approximate it with a finite number of parameters.

Among the various proposals, the most successful is the Fourier Neural Operator (FNO) [Li et al. (2021)]. By assuming a stationary kernel $\kappa(x,y) = \kappa(x-y)$, equation (17) reduces to a convolution integral. By applying the Fourier transform $\mathcal{F}$, the convolution becomes a product in the frequency domain. By applying an inverse Fourier transform $\mathcal{F}^{-1}$, the operation can be expressed as:

$$(\mathcal{K}_t(v_{t-1}))(x) = \mathcal{F}^{-1}(R(k;\theta)\mathcal{F}(v_{t-1}))(x) \tag{18}$$

where $R(k;\theta)$ is the target of the learning. By setting a maximum wavenumber $k$ to consider, FNO is capable of handling inputs at any resolution with the same model parameters. While satisfying discretization invariance, the use of the Fast Fourier Transform (FFT) enables both expressiveness and efficiency.

$\mathcal{P}$, $\mathcal{Q}$, $W_t$, and $\mathcal{K}_t$ are linear transformations with trainable parameters, while $\sigma_t$ is a predefined nonlinear activation function. The key point in the model construction is that most of the components are local operators, limiting the discussion of discretization invariance primarily to the integral operator $\mathcal{K}_t$. The FNO has become the de facto standard in operator learning. It is widely used in applications and serves as a baseline for the development of new models.

Note that the term neural operator (NO) can refer to two different concepts. In some contexts, it broadly refers to any operator learning model using NNs, including DeepONet. In other contexts, it refers more specifically to models with the structure discussed here, not including DeepONet. It is necessary to infer the meaning from the context. In this article, the latter, narrower definition will be used.

**4.3 Applications in Seismology**

NOs have been increasingly applied to seismic wave propagation. A simulation data-driven approach (Section 2.1) is used to learn the relationship between the velocity structure $a(x)$ and the wavefield $u(x)$. Yang, Y. et al. (2021) constructed an FNO for acoustic waves in two-dimensional structures, demonstrating the basic application of trained models, such as forward computations in full waveform inversion (FWI). Subsequently, Yang et al. (2023) extended it to elastic waves and Lehmann et al. (2024) addressed three-dimensional velocity structures.

Operator learning can directly infer solutions from phenomena (Figure 1). For example, by training on satellite observation data, a numerical weather prediction model with higher resolution and predictive accuracy compared to traditional methods has been developed [Pathak et al. (2022)]. However, such observational data-driven approaches (Section 2.1) require big data, which limits their applicability in seismology. One of the few applicable areas is seismic wave detection. Sun et al. (2023) proposed an NO model that performs phase picking from multiple observation stations, incorporating their spatial arrangement. Owing to discretization invariance, once the model is trained on a given observation network, it can be applied to any configuration.

**4.4 Methodological Advancements**

While FNOs have shown excellent performance in various tasks, a significant limitation is that the input and output spaces must basically be the same, which restricts its ability to represent general operators. In seismic wave propagation, if we want to model the spatiotemporal wavefield $u(t,x)$ from the spatial velocity structure $a(x)$, adjustments are necessary, such as introducing a dummy time variable for the velocity structure $a(t,x)$ and modeling the constant time evolution $u(t_0,x) \mapsto u(t_0 + \Delta t, x)$. Moreover, FFT is usually applied to a uniform grid in a rectangular domain, but in seismology, it is desirable to adjust the grid size near faults and surrounding areas, as well as to handle topography. There are active studies on extending NOs to handle more general geometry [Li et al., (2024a), Liu et al. (2024)].

Efforts are being made to improve FNOs for large-scale problems. Since the number of parameters considerably increases as the layers become deeper in the basic structure (equation 14 and Figure 6), the U-shaped neural operator was proposed where the number of discretization $m$ and the dimension of the latent space $d_v$ are progressively changed across layers [Rahman et al. (2023)]. This was used by Yang et al. (2023) in the previous section. Since the computational complexity of the FFT increases exponentially with the spatial dimension, the factorized Fourier neural operator was proposed where the FFT is performed separately for each spatial dimension [Tran et al. (2023)]. This was applied by Lehmann et al. (2024) to handle three-dimensional structures.

When solving the solution operator of known PDEs, physics-informed learning has been incorporated. Wang et al. (2021) proposed physics-informed DeepONet, where derivatives in the physics loss are computed using automatic differentiation, allowing the general solution of a PDE without training data. Li et al. (2024b) developed the physics-informed neural operator (PINO) based on FNO. A two-step training process was proposed: after learning the solution operator $G: \mathcal{A} \to \mathcal{U}$ using both data loss from numerical simulations and physics loss from PDEs, the solution $G(a) \in \mathcal{U}$ for a given condition $a \in \mathcal{A}$ is fine-tuned using only the physics loss. PINOs have technical challenges with automatic differentiation, and it is discussed to be efficient to apply the chain rule to analytical expressions of derivatives for each component in equation (14). Although some literature on PINOs claims that PINNs can only handle single solutions, PINNs can perform simultaneous solutions for finite-dimensional parameters, as discussed in Section 3.4.1. The primary advantage of PINOs is the ability to handle infinite-dimensional parameters (Table 2). However, PINNs provide continuous solutions, while PINOs provide discrete representations. Choosing the right method depends on the problem at hand.

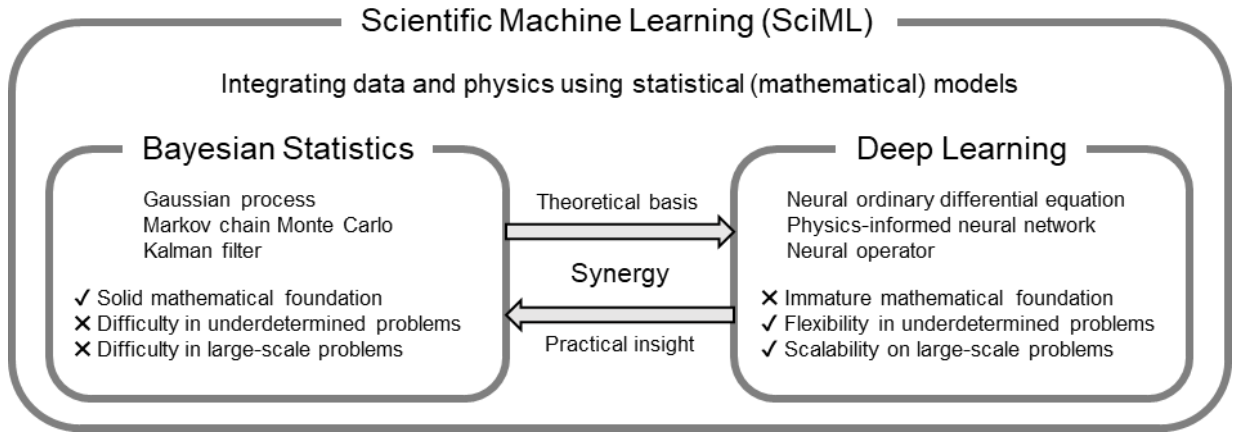

Figure 8. Broad definition of scientific machine learning (SciML). In addition to deep learning models (SciML in a narrow sense), conventional statistical and mathematical models such as Bayesian statistics can be regarded as SciML approaches.

Finally, while NOs were shown to satisfy discretization invariance, meaning they converge to the same operator in the limit of finer discretization [Kovachki et al. (2023)], it was pointed out that they do not necessarily satisfy continuous–discrete equivalence—the property that no aliasing errors occur at finite discretization [Bartolucci et al. (2024)]. Moreover, it is said that NOs, particularly PINO, have the property of superresolution—the ability to extrapolate to high-frequency components [Azizzadenesheli et al. (2024)], but the theoretical foundation is not necessarily clear. The basic structure of NOs (Figure 6) is not fixed, and there is potential for higher-performing model designs to emerge through continued theoretical and applied research.

**§5. Seismology Through the Lens of SciML**

In the previous sections, we reviewed the trends of SciML in seismology. On the other hand, even before the advent of SciML, seismology had already integrated physical models with observational data to carry out inference and prediction. This chapter reexamines traditional methods and deep learning through the lens of SciML, discussing their relationships and the potential for further development.

**5.1 Broad Definition of SciML**

Machine learning techniques, such as deep learning, have achieved significant success in human-centric fields like image recognition, natural language processing, and autonomous driving. As its application scope broadens, the need for domain-specific models in the natural sciences has been recognized, giving rise to the term scientific machine learning (SciML), increasingly regarded as a distinct area in machine learning [Baker et al. (2019)]. If we define SciML in broad terms, it can be summarized as "a research field that aims to enhance inference and prediction of natural phenomena by integrating observational data with physical models using statistical or mathematical models." This broad definition implies that SciML encompasses not only the latest deep learning-based methods (Figure 1) but also the sophisticated data analysis developed across various natural science disciplines, including seismology (Figure 8). Gaussian processes, Markov Chain Monte Carlo, and Kalman filters play central roles in inversion analysis and data assimilation in the Bayesian framework. More broadly, mathematical models such as stochastic processes and reduced-order models are also employed.

It is particularly interesting that seemingly different approaches have emerged to achieve a common goal. Bayesian statistics provides highly reliable inference with solid mathematical foundation. However, it often requires that governing laws be known and faces computational challenges for large-scale problems. In contrast, deep learning scales well to large problems and, being originally developed as a data-driven approach, can be applied even in cases where physical knowledge is incomplete (Section 3.4.2). However, the mathematical foundation is still underdeveloped, making it difficult to quantify confidence intervals or provide robust justifications, especially for predicting rare events. By leveraging the complementary strengths of these two approaches, a fruitful interplay can be developed. Traditional methods can provide deep learning with a theoretical basis, while deep learning can offer practical insights through empirical analysis, enhancing the methods on both sides (Figure 8). This synergy holds great potential for advancing our understanding of earthquake phenomena.

**5.2 Research Examples**

SciML defined broadly above is vast and diverse. This section focuses on the author's research using Bayesian statistics and mathematical models.

**5.2.1 Seismotectonics × Bayesian Inversion Analysis**

Understanding the regional-scale distribution of deformation and stress fields in the Earth's crust is fundamental to comprehending tectonic phenomena, including earthquakes and volcanic activity. Since observational data are typically limited to the near-surface and are sparse and noisy, uncertainty quantification is essential in estimating continuous fields. To address this, inversion methods that incorporate physical laws in the form of observation equations into Bayesian inference have been developed.

Okazaki et al. (2021a) proposed to apply a basis function expansion method [Yabuki and Matsu'ura (1992)] to estimate the strain rate field from GNSS velocity data. This approach enabled objective determination of the degree of smoothing and was quantitatively shown to produce a higher-resolution and more stable continuous field compared to the commonly used method by Shen et al. (1996). Application to the strain rate field around Japan clarified not only the Niigata-Kobe tectonic zone [Sagiya et al. (2000)] but also a low-strain region from northern Kanto to Aichi and spot-like high-strain regions that align with the volcanic distribution in northeastern Japan [Takada and Fukushima (2013)]. Strain rate maps for three different periods covering Japan have been produced using this method [Fukahata et al. (2022)] offering valuable references for studies on deformation and stress fields in the region.

Okazaki et al. (2022a) estimated the spatiotemporal variations of the stress field around Japan through inversion analysis of earthquake CMT data. Notably, they revealed that the temporal evolution of stress in the focal region of the 2011 Tohoku earthquake varied depending on the location. The basis function expansion method [Terakawa and Matsu'ura (2008)] faces computational difficulty when incorporating time, as the parameter space becomes four-dimensional. To overcome this, they introduced Gaussian processes [e.g., Rasmussen and Williams (2006)], a Bayesian inference whose computational complexity is independent of the dimensionality of the model. To apply Gaussian processes to inverse problems, they derived the relationship of covariances that the estimated field must satisfy. This approach has enabled the inversion of high-dimensional geophysical data that was previously difficult to analyze.

In summary, basis function expansion methods are more effective for low-dimensional, dense data with fewer model parameters. Gaussian processes are more effective for high-dimensional, sparse data with fewer observational data. It is crucial to choose the appropriate method for the specific problem.

**5.2.2 Strong Ground Motion × Optimal Transport**

Optimal transport theory is a mathematical framework that explores the geometry of probability spaces. Since the 2010s, its applications in machine learning have been growing [e.g., Peyré and Cuturi (2019)]. In particular, the Wasserstein distance, known for preserving the geometric structure of the original space, has been applied to loss functions in machine learning and a similarity measure of waveforms in seismology [Métivier et al. (2016)] .

Broadband ground motion prediction refers to the technique of forecasting seismic waveforms across a wide frequency range for scenario earthquakes. This is critical for designing structures and disaster prevention planning. A hybrid approach is commonly used, combining long-period waveforms from physics-based simulations with short-period waveforms generated through statistical methods. However, a challenge arises due to potential inconsistencies near the matching frequency, where the two independent methods are joined. Okazaki et al. (2021b) proposed a machine learning model that learns past ground motion records to transform simulated long-period waveforms into broadband waveforms consistently. Specifically, a nonlinear dimensionality reduction known as t-SNE [van der Maaten and Hinton (2008)] was used to construct a latent space where time characteristics are evaluated using the Wasserstein distance as a similarity measure. Then, an NN is used to learn frequency characteristics and both time and frequency features were combined to synthesize seismic waveforms. Aquib and Mai (2024) implemented a similar approach using FNOs.

The Wasserstein distance is applied to positive, normalized probability distributions, while seismic waveforms have both positive and negative values and varying amplitudes. Therefore, applying this metric to seismic waveforms requires appropriate transformations. Sambridge et al. (2022) proposed a method that treats waveforms as 2D images and calculates the Wasserstein distance by projecting them onto the time and amplitude axes. Okazaki and Ueda (2023) pointed out, through simple examples, that this approach can lead to degenerate distance. To address this issue, they showed that the sliced Wasserstein distance [Bonneel et al. (2015)]—which projects data in all directions, rather than just two—can resolve these degeneracies. While these approaches provide potential solutions, defining an optimal transport distance that is fully suited to seismic waveforms is still an area of ongoing research.

### 5.3 Outlook

SciML based on deep learning, especially physics-informed learning such as PINNs, is steadily advancing toward real-world applications in seismology. Section 3.4 discussed research directions that leverage the flexibility of deep learning (Figure 4). In this section, we will explore the trends and prospects of SciML research, focusing on two key challenges that have long been addressed by traditional methods: uncertainty quantification and large-scale modeling.

### 5.3.1 Uncertainty Quantification

In geophysics, observational data are typically localized, sparse, and noisy, so uncertainty quantification is critical in inversion analysis. Bayesian inference is commonly used for this purpose, as it yields probabilistic representation of estimation results (posterior distribution) by combining prior information on the physical system (prior distribution) with the fitting on observational data (likelihood). For linear models, such as basis function expansion and Gaussian processes, the assumption of Gaussian noises allows for the analytical calculation of the posterior distribution. However, approximate methods are required for Bayesian Neural Networks (BNNs), which is a Bayesian formulation of nonlinear NN models.

The integration of BNNs into PINNs leads to what is called Bayesian PINNs (B-PINNs). In a pioneering study by Yang, L. et al. (2021), two methods were compared: Hamiltonian Monte Carlo (HMC), a sampling method, and variational inference (VI), which approximates the posterior as a product of independent Gaussian distributions. The study concluded that HMC provided more accurate uncertainty quantification. However, the application of HMC-based B-PINNs has been mostly limited to low-dimensional problems like time-series predictions [Linka et al. (2022)]. The reason is that sampling from the vast parameter space of an NN using HMC comes with high computational costs, and it often fails to converge sufficiently in large-scale models.

To address the challenges, recent developments have focused on B-PINNs based on particle-based variational inference (ParVI). ParVI represents the posterior distribution as an empirical distribution of multiple particles (typically around 100) and optimizes their configuration to best approximate the true posterior. Compared to traditional variational inference, ParVI has higher approximation capabilities, and unlike sampling methods, it requires fewer particles and is parallelizable, making it well-suited for large-scale problems. Gou et al. (2023) applied Stein variational inference [Liu and Wang (2016)], a representative ParVI method, to traveltime tomography, demonstrating its effectiveness. Agata et al. (2023) pointed out that the posterior distribution of NN parameters tends to exhibit stronger multimodality compared to the posterior distribution of the estimated quantities. They developed a method using ParVI in function space [Wang et al. (2019)], which directly estimates the posterior distribution of the estimated quantities. This approach offers an additional advantage of making it easier to set physically meaningful prior distributions in the function space. Advancing such cutting-edge approximate inference methods is essential toward large-scale problem analysis.

### 5.3.2 Large-Scale Modeling

Megathrust earthquakes at plate boundaries are multi-scale and complex phenomena, causing strong ground motions, crustal deformation, and tsunamis through the processes of fault rupture and seismic wave propagation. To better understand and predict the wide-ranging impacts of such earthquakes, high-resolution, large-scale simulations that faithfully reproduce real-world rupture processes and subsurface structures are essential. Traditional numerical simulations, such as the finite element method, have played a key role in this effort. PINNs have advantages in incorporating observational data and the number of model parameters does not explicitly depend on the scale of the problem, which makes them suitable for data assimilation in large-scale problems.

Examples of applying PINNs to large-scale problems include simultaneous solutions for fast computation of traveltimes between any two locations in the regional or global velocity structure model [Smith et al. (2022), Taufik et al. (2023), Agata et al. (2024)]. Additionally, Agata et al. (2025) performed a sequential analysis, estimating the velocity structure from seismic exploration data and subsequently using the derived structure for hypocenter estimation, including uncertainty quantification using ParVI. This suggests the potential for efficiently calculating uncertainty propagation in complex seismic phenomena.

Seismic wave propagation is a multi-scale phenomenon, as it has influence on wide regions but is sensitive to local heterogeneities. A challenge for NNs is the spectral bias, where NNs tend to first learn long-wavelength components, making it difficult to improve the accuracy of short-wavelength components [Rahaman et al. (2019)]. Overcoming this optimization challenge is crucial for realizing large-scale modeling.

The application of NOs is also worth considering. However, obtaining sufficient observational data for seismic modeling is challenging, and the computational cost of generating a large amount of simulation data for large-scale problems is prohibitively high. In this context, the use of PINOs is a promising alternative. There is research on the simultaneous solution of traveltimes in arbitrary velocity structures relying solely on physics-informed learning [Song et al. (2024)]. For more complex phenomena, methods that derive solution operators from relatively small amounts of simulation data combined with physical laws are considered more promising (Section 4.4). Since operator learning enables rapid inference of solutions for arbitrary conditions, it also supports ensemble-based uncertainty quantification [Pathak et al. (2022)].

To realize large-scale modeling entirely through SciML approaches based deep learning, significant innovations in optimization methods that considerably reduce computational costs would be necessary. Rather than viewing SciML as a replacement for traditional numerical simulations, it may be more effective to see it as a complementary technology. Hybrid analysis methods that combine the strengths of numerical simulations, PINNs, and NOs—such as simulation data-learning in PINOs—may provide effective solutions.

§6. Conclusions

This article provided an overview of scientific machine learning (SciML), which uses machine learning, especially deep learning, to describe and predict natural phenomena, focusing on its fundamental concepts and research trends in seismology. A key feature of SciML is not merely applying general models developed in other fields to scientific problems, but rather integrating physical knowledge into the modeling process.

Physics-informed neural networks (PINNs) can perform forward analysis of differential equations by incorporating the residuals from physical laws into the loss function of NNs. When observational data are available, PINNs can also address inverse problems and data assimilation. In seismology, the application of PINNs has expanded from traveltime and wave propagation to crustal deformation and fault friction. This approach, known as physics-informed learning, is highly flexible. It has begun to yield intriguing results in areas where traditional methods struggle, such as simultaneous solutions in parameter spaces, the analysis of underdetermined systems, and regularization based on partial physical knowledge. PINNs have great potential for diverse research applications (Figure 4).

Neural operators (NOs) are methods for operator learning, aimed at modeling operators that describe relationships in infinite-dimensional spaces, typically in a data-driven manner. NOs are proposed to model natural phenomena, such as the general solutions of differential equations. Since NOs require extensive observational or simulation data, many studies in seismology focus on using them as surrogate models for seismic wave propagation. Methodological advancements for large-scale problems are actively being developed, and among these, PINOs—which combines operator learning with physics-informed learning—holds significant potential for future applications in SciML.

In general, SciML refers to a set of methods based on deep learning that have gained prominence around 2020. However, when considered more broadly as an approach that integrates observational data with physical models through statistical models, advanced Bayesian statistics and mathematical models have long been developed in seismology. The introduction of deep learning, which is more flexible and scalable than mathematically rigorous traditional methods, marks an interesting period where the complementary relationship between these two approaches has begun to be discussed. While the current trend mainly focuses on leveraging traditional methods to enhance deep learning, insights flowing in the reverse direction, as well as hybrid approaches that combine the strengths of both methods, are expected to create synergistic effects (Figure 8), which may open up new ways to understand earthquake related phenomena.

**Acknowledgments**

This article was written upon the recommendation of Dr. Yukitoshi Fukahata, the editor-in-chief of Zisin (Journal of the Seismological Society of Japan), following the receipt of the 2023 Young Researcher Encouragement Award by the Seismological Society of Japan. I appreciate Dr. Kazuro Hirahara, Dr. Hiroyuki Fujiwara, and Dr. Yukitoshi Fukahata for nominating me for the Award. I thank Dr. Yuji Itoh, the editor, and Dr. Masayuki Kano for valuable comments on the manuscript. I am grateful for Dr. Ryoichiro Agata for insightful discussions in preparing the manuscript.